\documentclass[aps,prb,twocolumn,10pt]{revtex4}
\usepackage{graphicx}
\usepackage{epsfig}
\usepackage{amsmath}
\usepackage{amssymb}
\usepackage{amsfonts}%
\begin{document} 
\title{Gutzwiller scheme for electrons and phonons: the half-filled Hubbard-Holstein model} 
\author{P. \ Barone$^1$, R. \ Raimondi$^2$, M.\  Capone$^{3,4}$, C. \ Castellani$^3$, and M. \ Fabrizio$^1$} 
\affiliation{$^1$International School for Advanced Studies (SISSA) and CNR-INFM-Democritos National Simulation Centre,
Via Beirut 2-4  34014 Trieste Italy}
\affiliation{$^2$CNISM and Dipartimento di Fisica ``E. Amaldi'' Universit\`a di Roma Tre, Via della Vasca Navale 84, 00146 Roma, Italy}
\affiliation{$^3$SMC, CNR-INFM and Dipartimento di Fisica, Universit\`a di Roma ``La Sapienza'', P.le Aldo Moro 2, 00185 Roma, Italy}
\affiliation{$^4$ ISC-CNR, Via dei Taurini 19, 00185 Roma, Italy}

\begin{abstract}

We analyze the ground-state properties of strongly-correlated electrons coupled with phonons by means of a generalized Gutzwiller wavefunction which includes phononic degrees of freedom. We study in detail the paramagnetic half-filled Hubbard-Holstein model, where the electron-electron interaction can lead to a Mott transition, and the electron-phonon coupling to a bipolaronic transition. We critically discuss the quality of the proposed wavefunction in describing the various transitions and crossovers that occur as a function of the parameters. Previous variational attempts are recovered as particular choices of the wavefunction, while keeping all the variational freedom allows to access regions of the phase diagram otherwise inaccessible within previous variational approaches.
\end{abstract} 

\date{\today} 
\maketitle 

\section{Introduction}

Understanding the interplay between electron-phonon (e-ph) and strong electron-electron (e-e) interactions can be important to understand the properties of many compounds, such as manganites\cite{imada} or fullerides\cite{gunnarson}.
This problem has recently received a renewed attention, after a number of experiments on high-T$_c$ superconducting cuprates testified for a non trivial role of the e-ph coupling in the underdoped region, where correlation effects are 
important and standard e-ph theories are hardly reliable\cite{schneider2001,lanzara2001,zhou,lee2006,gweon2007}.
Indeed a full theoretical understanding of lattice effects in strongly-correlated systems is still lacking when the two interactions have intermediate or large strength.

The essential features of a strongly-correlated system are captured in the celebrated Hubbard model\cite{hubbard} and in its extensions. On the other hand, polaron formation due to e-ph coupling has been extensively studied in the absence of e-e repulsion, mainly in the limit of low electron density within the molecular-crystal model first introduced by Holstein\cite{holstein}. Nonperturbative approaches such as Quantum Monte Carlo, Density Matrix Renormalization Group and Dynamical Mean-Field Theory (DMFT) have enormously contributed to the understanding of these models and have been successfully applied to the Hubbard-Holstein model\cite{huang2003,jeckelmann2006,dmft-koller,dmft-coreani,dmft-capone,dmft-paci}.
Nonetheless these methods require a considerable numerical effort, and it is not unfair to say that approximate analytic results, even if less accurate, are still highly desirable.

From this point of view, variational approaches are a very natural choice as they provide immediate information on the ground-state properties. In the field of strongly-correlated systems, the trial wavefunction introduced by Gutzwiller\cite{gutzwiller} and then generalized by B\"unemann and coworkers\cite{bunemann1998} has proved quite accurate for electronic systems with local interactions. The underlying recipe is to construct the correlated wavefunction starting from an adequate uncorrelated one and projecting out the energetically unfavored states. The equivalence with the Kotliar-Ruckenstein (KR) slave-boson saddle-point solution\cite{kotliar} allows also to go beyond the mean-field approximation in a controlled way.
For the Holstein model several trial wavefunctions have been proposed generalizing the Lang-Firsov transformation\cite{lang}, which yields the exact solution in the atomic limit, and it is believed to capture the essence of polaronic physics, i.e., the entanglement between electronic and phonic degrees of freedom. These kind of approaches have been successfully applied to the Holstein model, mainly in the limit of low carrier density or for spinless electrons\cite{emin,toyozawa, merrifield,zheng1988, feinberg1990, romero1998}. In fact, a further complication arises at finite density, since the e-ph coupling induces an effective e-e attractive interaction that cannot be handled in an independent particle picture. When the Hubbard repulsion is included, it directly competes with such phonon-mediated attraction.

The typical strategy to treat correlated fermions coupled with phonons consists in deriving an effective model for electrons by averaging the full Hamiltonian over a suitable trial phonon wavefunction (which also implies the neglect of any electron-multiphonon residual interaction)\cite{ihle1995}, and then to resort to some appropriate technique in order to tackle the electronic effective interaction. This can be done for instance by means of a Hartree-Fock decoupling scheme\cite{zheng1988} or exploiting the more accurate KR slave-boson mean field\cite{fehske,perroni,io_vlf} suitable for local e-e interactions.
As discussed in details in Refs. \onlinecite{feinberg1990,ranninger92} the assumption of a factorized wavefunction becomes questionable in the regime where retardation effects between the motion of electrons and lattice deformations imply a strong exchange of momentum.

In order to better describe the interplay between lattice vibrations and local electronic configurations,
we introduced in a previous paper\cite{io_gpw} a Gutzwiller Phonon Wavefunction (GPW) to account for the e-ph coupling besides strong electronic correlation and benchmarked it in the infinite-$U$ limit of the Hubbard-Holstein model by comparing with exact DMFT calculations. In this limit a finite phonon-driven attraction is obviously ineffective against the infinite repulsion, and the only effect of phonons is the formation of polarons.

In this paper we turn our attention to the finite-$U$ regime, analyzing in details the paramagnetic solution (i.e., neglecting broken-symmetry phases). Relaxing the infinite-$U$ assumption, the competition between repulsion and attraction becomes effective, leading to a rich phase diagram. We focus on the half-filling case, in which the e-e and e-ph interactions can drive, respectively, the Mott and bipolaronic metal-insulator transition, so that their interplay is both more transparent and more qualitative than for arbitrary fillings. Out of half-filling the system is indeed always metallic in the absence of symmetry breaking, and the effect of the competition is essentially quantitative and therefore less clear\cite{outofhalffilling}. 

The key property of our GPW wavefunction is to treat simultaneously electrons and phonons from the onset by introducing a generalized Gutzwiller projector where phonon quantum states are determined by the local electronic configurations rather than by the average electron density. This allows to address the modification induced by the electronic correlation onto the lattice groundstate. Therefore, even if the multiphonon adiabatic regime of the weakly-correlated system is probably still misrepresented, we expect to capture the
essential features of the model in the presence of a sizeable repulsive interaction, when the relevant physics is essentially local because the electron motion is slowed down by correlations.
We derive a set of variational equations for the phonon wavefunctions and show that standard approaches can be obtained as particular choices for the solution of such equations. We obtain a phase diagram which strongly depends on the adiabaticity regime, and critically discuss our results and the quality of the proposed wavefunction.

The paper is organized as follows. In Section \ref{sec:method} we introduce the model Hamiltonian and the relevant equations for the general case of arbitrary electron filling, and then we specialize them for the half-filling case we are interested in. At the end of this section we derive standard variational approaches for the Holstein problem as particular variational ansatzs on the Gutzwiller wavefunction. In Section \ref{sec:phasedia} we present the general solution provided by our method for the half-filled system, whose properties are analyzed in the following Section \ref{sec:polform}. We then discuss in Section \ref{sec:discussion} the quality of the proposed wavefunction by comparing it with other variational wavefunctions.

\section{Model and method}\label{sec:method}

The Hamiltonian of the Hubbard-Holstein model at half-filling reads:
\begin{eqnarray}\label{ham_hh}
H&=&-t\sum_{<i,j>,\sigma} c^\dagger_{i\sigma} c_{j\sigma} + \frac{U}{2}\sum_i (n_{i}-1)^2 +\omega_0\sum_i a^\dagger_i a_i \nonumber \\&&  +
\alpha\omega_0\sum_i (n_i-1)(a_i^\dagger+a_i),
\end{eqnarray}
where $c_{i\sigma}\; (c^\dagger_{i\sigma})$ and $a_{i\sigma}\;
(a^\dagger_{i\sigma})$ are respectively annihilation (creation) operators for
tight-binding electrons with spin $\sigma$ and for optical phonons of 
frequency $\omega_0$ on site $i$, $t$ is the nearest-neighbor
hopping amplitude, $U$ the local Hubbard repulsion and $\alpha$ parameterizes 
the coupling between  local displacements and electronic density fluctuations $n_i-1$. Both interaction terms have been written in a particle-hole symmetric form which enforces the half-filling condition.
We introduce an adiabaticity parameter defined as $\gamma=\omega_0/D$, measuring the ratio between the phonon and electron characteristic energies, and a dimensionless parameter $\lambda=2\alpha^2\omega_0/ D=2\alpha^2 \gamma$ which measures the strength of the e-ph coupling.  Here $D$ is the half bandwidth, which depends on the hopping parameter $t$ once a particular lattice is chosen.
We perform our calculations in an infinite-coordination Bethe lattice with semicircular density of states of half-bandwidth $D=2t$.\cite{notabethe} For this lattice, as in any infinite-coordination lattice, the expectation value of the Hamiltonian on the Gutzwiller state can be performed without further approximations, because all the configurations with the same average occupation number equally contribute to the quantum averages. The same approach can be used also for finite-dimensional lattices, where it represents a further approximation, usually called Gutzwiller approximation.
Another advantage of considering the infinite-coordination Bethe lattice is the possibility to compare with the exact results for this lattice, obtained through DMFT.\cite{dmft-koller,dmft-coreani,dmft-capone,dmft-paci}

We introduce the GPW as\cite{io_gpw}
\begin{equation}\label{gutzwiller}
\vert\Psi \rangle=\Pi_i {\cal P}_i (x_i)\vert\Psi_0\rangle,
\end{equation}
where $|\Psi_0\rangle$ is a Slater determinant, that we take as the non-interacting Fermi sea in order to describe metallic states. ${\cal P}_i (x_i)$ is a generalized Gutzwiller projection operator 
\begin{equation}
{\cal P}_i (x_i)=\sum_{\nu=0,1,2}\sqrt{\frac{P_\nu}{P^{(0)}_\nu}}\phi_{\nu} (x_i)|\nu_i\rangle\langle \nu_i|.
\end{equation}
Here $\vert \nu_i\rangle\langle \nu_i\vert $ are the projectors associated to the different local electronic states, empty site ($\nu =0$), singly occupied site ($\nu =1$) and doubly occupied site ($\nu =2$)ß. 
Our wavefunction therefore associates to each local electronic state a normalized {\sl first quantization} phonon wave function $\phi_\nu (x_i)$  depending on the displacement coordinate $x_i =(a_i +a_i^\dagger)/\sqrt{2}$, which has to be determined variationally, exactly like the probabilities of each of the local states $P_\nu$ ($P^{(0)}_\nu$ are the same quantities for the uncorrelated system, {\sl i.e.}, 
$P^{(0)}_0=(1-n/2)^2$, $P^{(0)}_1= n/(1-n/2)$, and $P^{(0)}_2=(n/2)^2$, $n$
being the average density per site). By imposing the standard constraints\cite{bunemann1998,attaccalite2003}
\begin{eqnarray}
\int dx_i \,\langle\Psi_0 \vert{\cal P}_i \vert^2|\Psi_0\rangle&=&1 \label{const:1},\\
\int dx_i\, \langle\Psi_0 \vert n_i \vert {\cal P}_i \vert^2\vert\Psi_0\rangle&=&n,\label{const:2}
\end{eqnarray}
and introducing the parameter $d=(P_0+P_2)/2$ and the doping $\delta =1-n$, one gets
$P_{0}= d+\frac{\delta}{2},\,
P_1=  1-2d \,\mbox{and}\,
P_2=  d-\frac{\delta}{2}
$
for the correlated occupation probabilities. Setting $\delta=0$, the three amplitudes are controlled by a single variational parameter, $d$. 
The variational energy per site is
\begin{eqnarray}\label{energy}
E&=&\sum_{\nu=0,1,2}P_\nu \langle h_0 (x)\rangle_\nu +\sqrt{2}\alpha\omega_0 \left[ P_0\langle x\rangle_0 -P_2\langle x\rangle_2\right]\nonumber \\
&-&2\vert\varepsilon\vert|S|^2+\frac{U}{2} \ (P_2+P_0)
\end{eqnarray}
where $h_0 (x)=(\omega_0/2)(-\partial_x^2+x^2))$ and
\begin{equation}
\langle O \rangle_\nu \equiv \int_{-\infty}^{\infty}{\rm d}x \  \phi^*_\nu(x) O \phi_\nu(x)
\end{equation}
indicates the average over the phonon wave function $\phi_\nu (x)$ of the operator $O$.
The second line of Eq.(\ref{energy}) represents the standard energy of the Gutzwiller approximation for the
Hubbard model with
\begin{eqnarray}
2\vert\varepsilon\vert=\,t\,\langle\Psi_0\vert \sum_{<i,j>,\sigma} c^\dagger_{i\sigma} c_{j\sigma} \vert\Psi_0\rangle
\end{eqnarray}
the free electron kinetic energy and
\begin{equation}\label{hopping_s}
S=\sum_{\nu=0,1} \sqrt{\frac{P_\nu P_{\nu+1}}{1-\delta^2}}\int {\rm d}x \ \phi^*_{\nu+1}(x)\phi_\nu(x)
\end{equation}
the renormalization factor that accounts for the reduced electron mobility
in the presence of e-ph and e-e interactions.
Then we have to minimize with respect to $d$ and the $\phi_{\nu}(x)$. The first minimization gives 
\begin{eqnarray}\label{mf_d}
&&U+( \langle h_0(x) \rangle_0+   \langle h_0(x) \rangle_2 -2 \langle h_0 (x)\rangle_1 )\nonumber\\
&&+\sqrt{2}\alpha\omega_0
(\langle x \rangle_0-  \langle x \rangle_2  )
-2|\varepsilon|\frac{\partial |S|^2}{\partial d}=0,
\end{eqnarray}
while the second yields the following non-linear second-order differential equations
\begin{eqnarray}
\frac{\epsilon_0}{P_0} \phi_0& =h_0(x)\phi_0 +\sqrt{2}\alpha\omega_0 x\phi_0
-\frac{2|\varepsilon|}{\sqrt{1-\delta^2}}S\sqrt{\frac{P_1}{P_0}}\phi_1  \label{mf_0}&\\
\frac{\epsilon_1}{P_1} \phi_1 &=h_0(x)\phi_1 
-\frac{2|\varepsilon|}{\sqrt{1-\delta^2}}\left( S^*\sqrt{\frac{P_0}{P_1}}\phi_0+S\sqrt{\frac{P_2}{P_1}}\phi_2 \right)\label{mf_1}&\\
\frac{\epsilon_2}{P_2} \phi_2 &=h_0(x)\phi_2 -\sqrt{2}\alpha\omega_0 x\phi_2
-\frac{2|\varepsilon|}{\sqrt{1-\delta^2}}S^*\sqrt{\frac{P_1}{P_2}}\phi_1 \label{mf_2}&
\end{eqnarray}
where the $\epsilon_\nu$'s are Lagrange multipliers enforcing the normalization conditions on the $\phi_\nu$'s. 

Before discussing in some detail the solution of Eqs.(\ref{mf_0})-(\ref{mf_2}), we briefly analyze their structure. In the absence of e-ph coupling they reduce to three equivalent Schr\"odinger equations for a free harmonic oscillator centered at each site independent on the electron occupancy. 
Therefore the phonon wavefunction is the standard gaussian, contributing $\omega_0/2$ to the on-site energy. Switching on the e-ph interaction has two major effects. The first  is to induce a shift proportional to $\pm\sqrt{2}\alpha$ to the phonon wavefunctions associated to the $\nu=0,2$ electron states, which is the expected effect in the atomic limit with the chosen form of the e-ph interaction; the second is a non-local coupling between wavefunctions related to different charge configurations, which appears to be proportional to the average kinetic energy of the electrons. This is not surprising, since one expects the electron dynamics to affect the phononic properties driving them away from the atomic limit scenario. Nonetheless, it is readily seen that this coupling term is inversely proportional to the adiabaticity ratio, that means, as it is well-known, that in the antiadiabatic regime, i.e. when phonons move faster than electrons, the interplay of phonon and electron dynamics is negligible: the lattice rearranges itself almost instantaneously with respect to slowly moving electrons and a charge-dependent shift of the phonons is sufficient to capture the ground-state properties of the system. On the other hand the coupling term becomes dominant in the opposite limit, i.e. for $\gamma\ll 1$; in this case each projected wavefunction depends heavily on the other wavefunctions and on the electron probability distribution, thus suggesting non trivial dependence on correlation effects.

At half-filling we have $P_0=P_2=d$ and $P_1=1-2d$, that imply $\phi_1(-x)=\phi_1(x)$ and $\phi_2(-x)=\phi_0(x)$ for the lowest-energy state. Therefore we are left with two variational equations for the lattice ground state coupled with Eq. (\ref{mf_d}). The latter can be recast in the more transparent form
\begin{eqnarray}\label{hf_d}
4d=1-\bar{u},
\end{eqnarray}
which depends from the important quantity
\begin{eqnarray}\label{baru}
\bar{u} = \frac{U - 2\,\sqrt{2}\,\alpha\omega_0\,\langle x\rangle_2 -2\, \langle h_0 (x)\rangle_1+2\,\langle h_0(x)
\rangle_2}{8\,\vert\varepsilon_0\vert\vert\langle\phi_2\vert\phi_1\rangle\vert^2\,}
\end{eqnarray}
measuring the effective degree of electronic correlation as the ratio between renormalized $U_{eff}$ (numerator) and phonon-renormalized kinetic energy $\varepsilon_{eff}$ (denominator)\cite{io_vlf}.

In spite of these simplifications, an analytical solution of our equations is still a difficult task. 
In the next section we present the general solution of Eqs. (\ref{mf_0})-(\ref{mf_2}). Here we briefly discuss some specific choices for the phonon wavefunction that allow for an analytical solution and reproduce popular variational approaches.

\paragraph{Gaussian ansatz for the phonon wavefunctions.}

As discussed above, in the atomic limit the phonon wavefunctions reduce to displaced harmonic oscillators
whose displacement $x_0$ depends on the local electron occupation according to $x_0(\nu)=\sqrt{2}\alpha (\nu-1)$. We can assume:
\begin{eqnarray}
\vert\phi_\nu\rangle=e^{i\sqrt{2}\alpha\,f_\nu \hat{p}}\vert 0\rangle
\end{eqnarray} 
where $\vert 0\rangle$ is the ground state of an undisplaced harmonic oscillator, $\hat{p}=-i(a-a^\dagger)/\sqrt{2}$ is the conjugate
momentum and $f_\nu$ is a parameter to be variationally determined.
Exploiting the particle-hole symmetry we can impose $f_2=-f_0=f$ and $f_1=0$ and obtain the variational ground-state energy per-site:
\begin{eqnarray}
E_0 = \frac{\omega_0}{2} \,- \,8d(1-2d)\,e^{-\alpha^2 f^2}\,\vert\varepsilon_0\vert \,+ && \nonumber \\
d[U - 2\alpha^2\omega_0\,f\,(2-f)], &&\label{vlf_energy}
\end{eqnarray}
that corresponds, within a constant, to the result of KR slave-bosons supplemented by the variational Lang-Firsov transformation (VLF)\cite{perroni,io_vlf}. This is not surprising, since the Gutzwiller approach is known to be equivalent to the slave-boson mean-field approach and the VLF describes the phonon wavefunctions as displaced Gaussians.

The mean-field solution is then determined by minimizing (\ref{vlf_energy}), that amounts to solve Eq.(\ref{hf_d}) with $\bar{u}=\exp(\alpha^2\,f^2)\,[U-2\alpha^2\omega_0\,f\,(2-f)]/8\vert\varepsilon_0\vert$
together with\cite{io_vlf}
\begin{eqnarray}\label{vlf_f_mf}
&&f=\left[1+\frac{2\vert\varepsilon_0\vert}{\omega_0}(1+\bar{u})\, e^{-\alpha^2f^2}\right]^{-1}.
\end{eqnarray}

\paragraph{Squeezed phonon state ansatz for the phonon wavefunctions.}

As widely discussed for the Holstein model\cite{zheng1988,fehske,feinberg1990,ihle1995}, the harmonic (gaussian) ansatz is not expected to be accurate except for the so-called light polaron case, that is realized when anharmonic fluctuations of the lattice are not so important, i.e. when $\lambda,\gamma\ll 1$ or when $\gamma\gg 1,\vert U_{eff}\vert$\cite{ihle1995}. Such anharmonic fluctuations can be partially captured by assuming that phonon wavefunctions are ``squeezed'' two-phonon coherent states (Variational SQueezed, VSQ). The simplest choice that is usually made is:
\begin{eqnarray}\label{ansatz:vsq}
\vert\phi_\nu\rangle=e^{-\alpha\,f_\nu (a^\dagger-a)}\,e^{-F(a a -a^\dagger a^\dagger )}\vert 0\rangle,
\end{eqnarray} 
where the e-ph induced displacement of the phonon field is still associated to the local charge state whereas the variational parameter $F>0$ controlling the squeezing of the wavefunctions is charge-independent, i.e., as noticed first in Ref. \onlinecite{feinberg1990}, it describes a kind of mean-field effect on the lattice due to the electron motion.
The ground-state energy computed over these phonon wavefunctions reads:
\begin{eqnarray}\label{sq_energy}
E_0 = \frac{\omega_0}{4}(\tau^2+\tau^{-2}) - 8d(1-2d)\,e^{-\alpha^2 f^2 \tau^2}\,\vert\varepsilon_0\vert + &&\nonumber\\ 
d[U - 2\alpha^2\omega_0\,f\,(2-f)],&&
\end{eqnarray}
where we have introduced $\tau^2=\exp(-4F)$. This is what one would obtain by applying the slave-boson mean field to the
effective polaron model derived by averaging over squeezed phonon states\cite{zheng1988,ihle1995}.
The mean-field equations are modified as follows:
\begin{eqnarray}
&&f=\left[1+\frac{2\vert\varepsilon_0\vert}{\omega_0}\,\tau^2\,(1+\bar{u})\, e^{-\alpha^2f^2\tau^2}\right]^{-1},\label{sq_f_mf}\\
&&\tau=\left[1+\frac{4\vert\varepsilon_0\vert}{\omega_0}\,\alpha^2\,f^2\, (1-\bar{u}^2) e^{-\alpha^2f^2\tau^2}\right]^{-1/4},\label{sq_t_mf}
\end{eqnarray}
with $\bar{u}=\exp(\alpha^2 f^2 \tau^2)\,[U-2\alpha^2\omega_0 f (2-f)]/8\, \vert\varepsilon_0\vert$. While $f$ still accounts for the effective e-ph-induced displacement of the ions, the variational parameter $\tau$ can be viewed as a measure of anharmonic fluctuations of the phonon wavefunctions.

\section{Half-filling phase diagram.}\label{sec:phasedia}

As we briefly discussed in the introduction, the paramagnetic phase of the half-filled Hubbard-Holstein model is an interesting test-field for our trial wavefunction since the interplay of the two local interaction mechanisms is more effective and at the same time more transparent because both the interaction terms are able to drive metal-insulator transitions of different nature. From a technical point of view, imposing the particle-hole symmetry simplifies the calculation, eliminating of the three self-consistency equations for the phonon wavefunctions. 
Finally, the half-filled phase diagram has been extensively studied by means of DMFT, providing us with an exact benchmark for the infinite coordination Bethe lattice.\cite{dmft-koller,dmft-coreani,dmft-capone,dmft-paci}

In this section we mainly focus on the metal-insulator transitions driven by the two interaction terms. In the absence of e-ph coupling a transition from a metal to a Mott insulator (MI) occurs at a critical $U_c$, whereas the half-filled Holstein model displays with increasing $\lambda$ a transition to a pair (bipolaron) insulator (BPI), the occurrence of such instabilities depending on the adiabaticity
regime\cite{caponeciuchi}.

In the present framework, a key parameter to characterize the properties of the system is the effective correlation parameter $\bar{u}$. From Eq. (\ref{hf_d}), $\bar{u}= 1$ corresponds to $d=0$, and $\bar{u}= -1$ to $d=1/2$.
These are sufficient conditions for, respectively, the Mott and bipolaronic transitions.
In fact, in the present mean-field description, the vanishing of the average double occupation signals the transition to a MI, with one electron per site, whereas $d=1/2$ corresponds to a system of electron pairs stuck to lattice sites, i.e., a BPI.
Once the existence of the insulating solution is proved, their thermodynamic stability must however be checked by comparing their ground-state energies (i.e., $E_0(MI)=\omega_0/2$ and $E_0(BPI)=[\omega_0+U-2\alpha^2\omega_0]/2$) with that of other possible (metallic) solutions. 
To gain a first insight about the combined role of the two interaction terms, we notice that the effect of phonons on the correlation parameter $\bar{u}$ is twofold. On one hand the e-ph interaction reduces the numerator of $\bar{u}$, but it also reduces the denominator normalizing the kinetic energy. This means that under different circumstances the two interacting mechanisms can either cooperate or compete in localizing the particles.

A full solution of the Gutzwiller equations can be easily achieved numerically by expanding the phonon wave functions in the complete basis of the eigenstates $|n\rangle$ of $h_0$, the harmonic oscillator centered around $x=0$, with eigenvalues ${\cal E}_n=\omega_0(n+1/2)$

\begin{equation}\label{exp}
\phi_\nu(x) =\sum_{n=0}^{\infty}c^{(\nu)}_n \langle x|n\rangle 
\end{equation}

This expansion is completely general and it does not introduce any further constraint or approximation.
Exploiting the particle-hole symmetry $c^{(0)}_n=(-1)^n c^{(2)}_n$ and plugging (\ref{exp}) in the GPW equations (\ref{mf_0})-(\ref{mf_2}) one gets:
\begin{eqnarray}\label{mf_ph-coeffs}
\left[\left(\frac{\epsilon_1}{1-2d} -{\cal E}_n\right)\delta_{nn'}
+8|\varepsilon_0| \, d\, c^{(2)}_n\,{c^{(2)}_{n'}}^* \right]\,c^{(1)}_{n'}&=&0 \nonumber\\
\left[\left(\frac{\epsilon_2}{d}-{\cal E}_n \right)\delta_{nn'}
+4|\varepsilon_0|\,(1-2d)\,c^{(1)}_n\,{c^{(1)}_{n'}}^*
\right. \hspace{0.8cm} && \nonumber \\
\left. -\sqrt{2}\alpha\omega_0\, x_{n'n}\right]c^{(2)}_{n'}&=&0\nonumber
\end{eqnarray}
which can be solved iteratively by keeping an arbitrary number of harmonic oscillator levels (in practice 
with 50-70 levels we get already very accurate results for the ground state); of course
their solution depends on the value of the parameter $d$, which has to
be determined self-consistently through Eqs. (\ref{hf_d}),(\ref{baru}) at each iteration.

\begin{figure}[h]
\includegraphics[width=8.5cm]{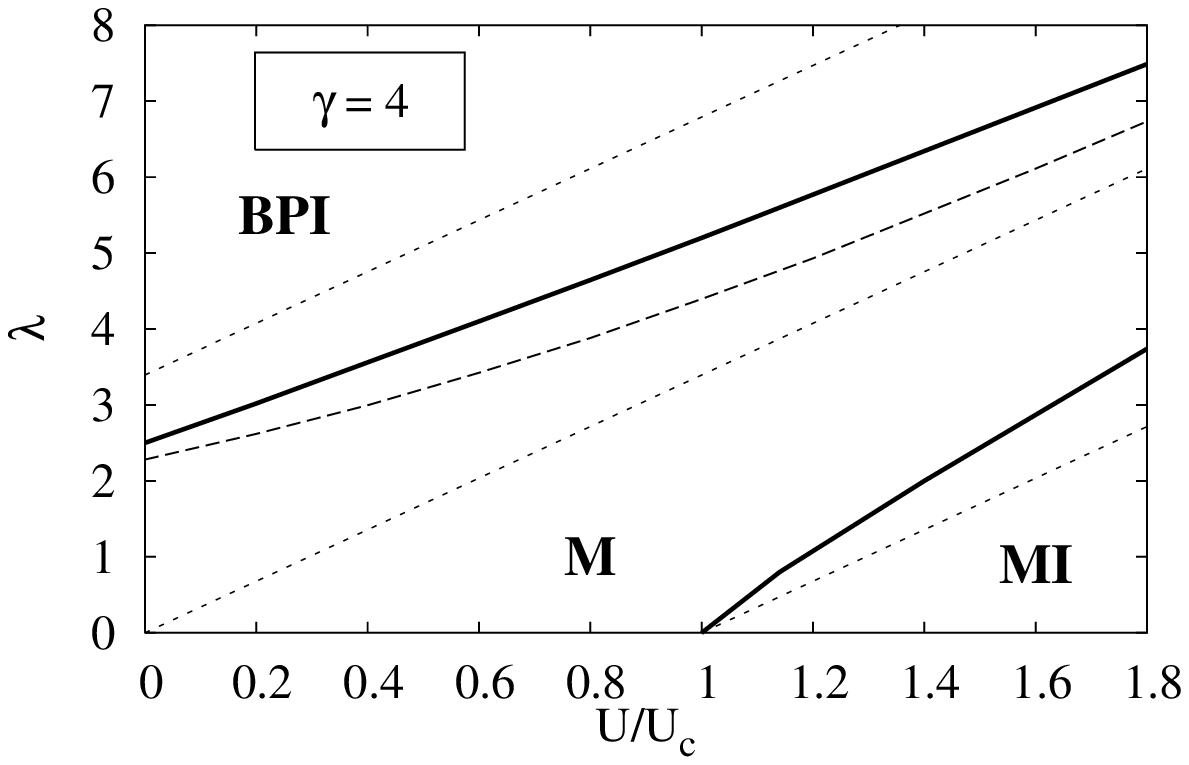}
\includegraphics[width=8.5cm]{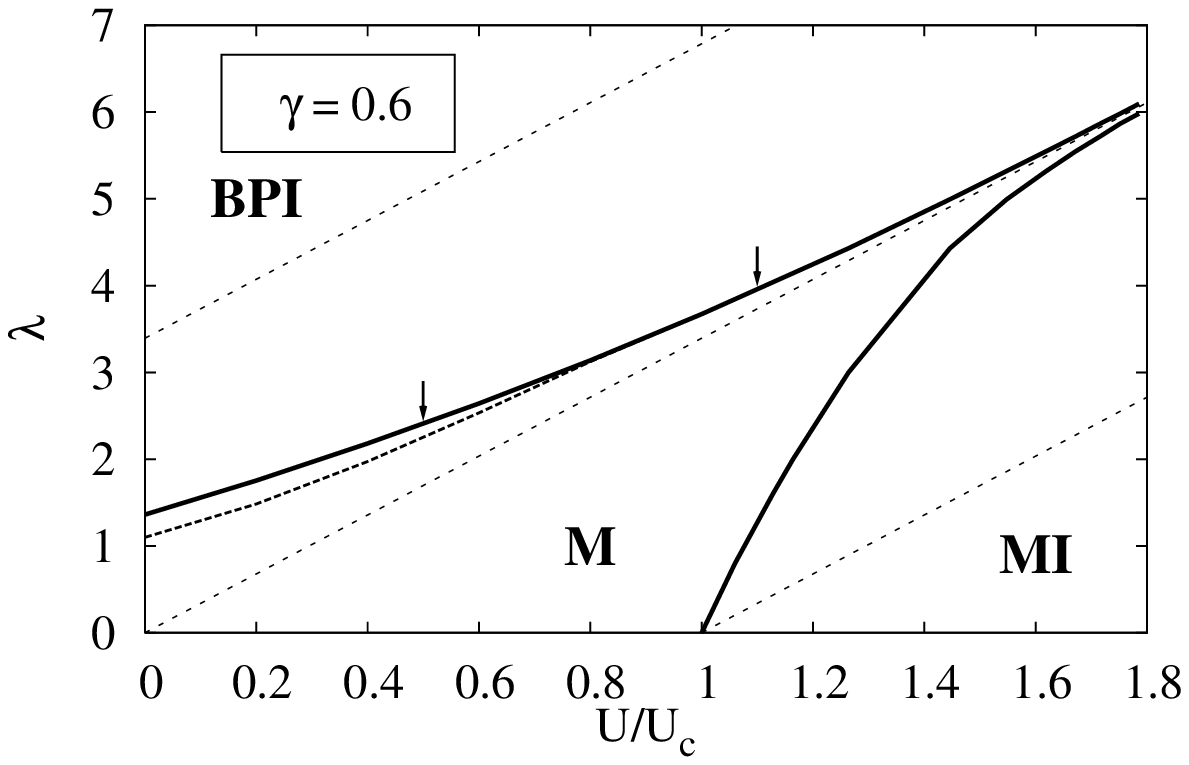}
\includegraphics[width=8.5cm]{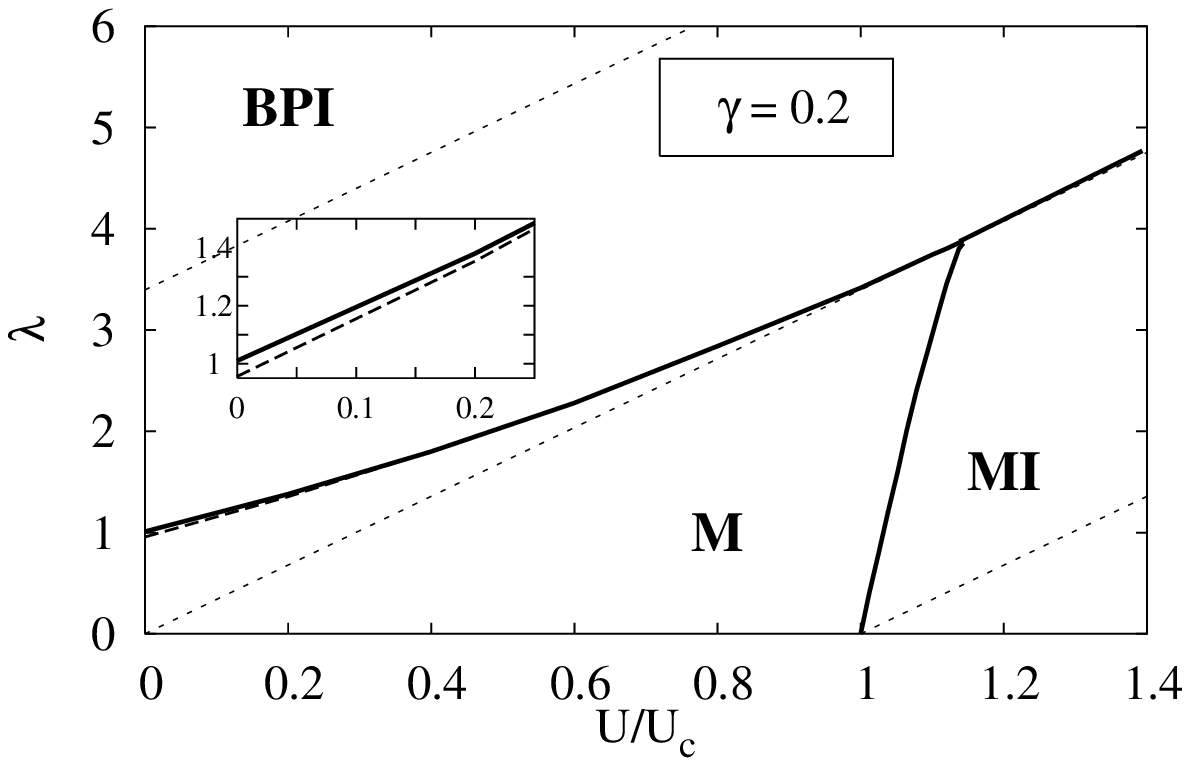}
\caption{Phase diagram of the half-filled Hubbard-Holstein model in the $U-\lambda$ plane as obtained in GPW for three different values of $\gamma$. Solid lines represent the metal-insulators transitions, dashed lines are the bipolaron formation lines obtained by imposing condition Eq.(\ref{polf_cond}) as discussed in Section \ref{sec:polform} (the inset shows the narrow region where a bipolaronic metal is found for $\gamma=0.2$). Thin dotted lines are the antiadiabatic ($\gamma\to\infty$) transition lines symmetric to the $U=\lambda D$ line, also displayed.  The vertical arrows in the middle panel mark the boundary of the region in which the transition to the BPI is of first order.
}\label{fig:phasedia}
\end{figure}

We show in Fig. \ref{fig:phasedia} the obtained phase diagram in the $U-\lambda$ plane, displaying three different phases, the two insulators MI and BPI and a correlated metal, which, as we will discuss in the next section, may show bipolaronic features before the metal-insulator transition. Since one expects the properties of the system to be strongly dependent on $\gamma$, as it happens in the absence of correlation, the phase diagram has been determined for three
different values of the adiabaticity parameter, namely $\gamma=0.2, 0.6, 4$.

In the antiadiabatic limit, where the phonon wavefunctions are simply displaced harmonic oscillators, one finds that
$\vert\langle\phi_1|\phi_2\rangle\vert=\exp(-\alpha^2/2)\approx 1$ and $\bar{u}\simeq [U-\lambda D]/U_c = U_{eff}/U_c$. The Holstein attraction becomes instantaneous and the system behaves essentially as a repulsive Hubbard model as long as $U_{eff}>0$ and as an attractive one for negative $U_{eff}$. On the repulsive side we will have a Mott transition, while on the attractive side the normal phase displays a metal-insulator transition between a metal and an insulating phase formed by incoherent pairs, which is the antiadiabatic limit of the BPI\cite{pairingmit}. Therefore transitions to insulating phases are second order and occurs symmetrically with respect to the $U_{eff}=0$ line, i.e. $U_{MI}=U_c+\lambda D$ and $\lambda_{BPI}D=U_c+U$ (thin dotted lines in Fig. \ref{fig:phasedia}).

Considering finite values of $\gamma$ we find that the main effect onto the Mott transition with respect to $\gamma\to\infty$ is to change the slope of the transition line at small $\lambda$, which remains second-order and is shifted to lower values of $U$ with decreasing $\gamma$. Increasing $\lambda$ the metal-insulator line bends towards the bisector $U=\lambda D$ when $\gamma$ is decreased. We notice that for small $\lambda$ the Mott transition line basically coincides with the prediction of the simple VLF approach. This is easily understood observing that, as long as the transition is of second order, it is defined by the condition $\bar{u}=1$. 
According to Eq. (\ref{sq_t_mf}), $\tau^2\rightarrow 1$ as $\bar{u}\rightarrow 1$, which means that squeezing effects do not improve the VLF  variational ansatz. Plugging the VLF expression for the phonon wavefunctions in Eq. (\ref{baru}) and imposing $\bar{u}=1$ one gets for small $\lambda$ the explicit relations:
\begin{eqnarray}\label{mott}
\lambda_{MI} D &=& \frac{U_c}{2\omega_0} (U-U_c) \hspace{2cm} \gamma\ll 1,\\
\lambda_{MI} D &=& \left[1-\frac{U}{2\omega_0}\right]^{-1} (U-U_c) \hspace{0.5cm} \gamma\gg 1,
\end{eqnarray}
in excellent agreement with previous DMFT phase diagrams\cite{dmft-koller,dmft-coreani,dmft-capone}.

Turning to the region of the parameter space dominated by the e-ph coupling, a more complex, strongly $\gamma-$dependent picture emerges. 
As long as $\gamma\gg1$, the metal-BPI transition is found to be second-order, and the condition $\bar{u}=-1$ is always satisfied at the transition. Since in this limit the $\phi_\nu$'s are essentially displaced harmonic oscillators, one finds:
\begin{eqnarray}\label{bpi}
\lambda_{BPI} D = U_c \left[ e^{-\frac{\lambda}{2\gamma}} + U/U_c \right].
\end{eqnarray}
As expected, the presence of $U$ obstacles the stabilization of the BPI and pushes the pair transition to higher values of $\lambda$. The ability of $U$ to compensate the phonon-mediated attraction is maximum for large $\gamma$, so that the same transition line moves to smaller $\lambda$ (yet larger than in the absence of $U$) as $\gamma$ is reduced. Decreasing $\gamma$ the scenario becomes more involved, and the order of the transition turns out to depend on both $\gamma$ and $U$. For $0.5 \lesssim \gamma \lesssim 1$ there is a window of intermediate values of $U$ in which the transition becomes of first order. Therefore the transition to the BPI is second-order at small values of $U$, turns to first order at a given $U$ depending on the considered $\gamma$ and then, at larger $U$, the transition becomes second-order again (e.g. at $\gamma=0.6$ we obtain a first-order transition for $U/U_c$ between 0.5 and 1.1). The metallic region comprised between the MI and the BPI appears to close asymptotically on the line $U_{eff}=0$ with increasing interactions strength. Eventually, for $\gamma \lesssim 0.5$ the transition to the BPI is first order for all values of $U$.

Some comments are in order about the evolution of the order of the transition to the BPI. For zero electron-electron interaction, DMFT shows a continuous transition\cite{caponeciuchi}. Thus the first-order nature of the metal-BPI transition at zero and small $U$ is due to the limitations of our variational approach.
Indeed more restrictive variational choices, like the VLF and the VSQ predict a first-order transition for $\gamma \lesssim \gamma_{1}$ (for the VSQ the limiting $\gamma_1$ is only slightly smaller than for VLF). 
The full solution of our GPW equations lowers the value of $\gamma_1$, but a more sophisticated wavefunction, able to capture nonlocal character of the ground state of e-ph coupled system, is required in adiabatic regimes, when retardation effects between the motion of electrons and lattice distortion of the lattice become particularly relevant.

 On the other hand a change in the order of the transition, from second to first one with increasing $U$, has been reported within exact DMFT at $U/U_c\simeq0.5$ for $\gamma=0.1$\cite{dmft-koller}, and it was associated in Ref. \onlinecite{dmft-paci} to an abrupt change in the electronic configuration from the Mott state where all the sites are singly occupied to the BPI where all the sites are either empty or doubly occupied. Therefore our GPW is able to correctly reproduce this highly non trivial evolution of the order of the transition as a function of $U$. We underline that more limited variational choices such as VLF and VSQ always find a first-order transition regardless the value of $U$. This result witnesses that the improvement brought by the GPW is particularly important in the presence of correlation effects.

Our analysis shows a remarkable agreement between our phase diagram and previous DMFT results for $\gamma=0.2$. Such an accuracy can only increase for larger $\gamma$, where we observe, e.g., a metallic phase which intrudes between the two insulators for large $U$ and $\lambda$. Unfortunately to our knowledge there are no detailed DMFT studies of the phase diagram for larger values of $\gamma$ to compare with our variational predictions.
 Recently, the phase diagram of the one-dimensional Hubbard-Holstein model at different values of the adiabaticity parameter has been obtained, and an intermediate metallic phase has been observed getting larger between the two insulating phases as $\gamma$ increases\cite{hardikar2007}, in qualitative agreement with our results.

\section{Quasiparticle weight, phonon displacement and bipolaron formation}\label{sec:polform}

Beside the metal-insulator transition associated to bipolaronic binding, a strong electron-phonon coupling can drive the formation of polaronic and bipolaronic states. A polaron is an electron strongly entangled with phonons, thus moving with a significantly enhanced effective mass. The residual interaction between polarons is attractive, leading to the formation of pairs, called bipolarons.
Strong Coulomb repulsion or large lattice fluctuations can avoid the metal-insulator transition due to bipolaron localization, and they can lead to a poorly metallic state of bipolarons. 
It has been repeatedly shown that the formation of polarons and bipolarons in the absence of electron-electron interaction occurs through a continuous crossover. In the following we introduce and discuss two quantities that allow us to characterize the (bi)polaronic character of the correlated metallic phase from the point of view of both electronic and phononic observables.

The metallic or insulating character of the variational solutions can be addressed by computing the quasiparticle renormalization factor, which corresponds to the inverse effective mass $m^{*}$ in the Gutzwiller approach:
\begin{equation}\label{z}
Z = \frac{m}{m^*} = 8d(1-2d)|\langle\phi_1|\phi_2\rangle|^{2} = (1-\bar{u}^2)|\langle\phi_1|\phi_2\rangle|^{2},
\end{equation}
that is zero when the system is a MI or a BPI and a nonvanishing quantity $\leq 1$ for metallic
solutions.
\begin{figure}[h]
\includegraphics[width=4.2cm]{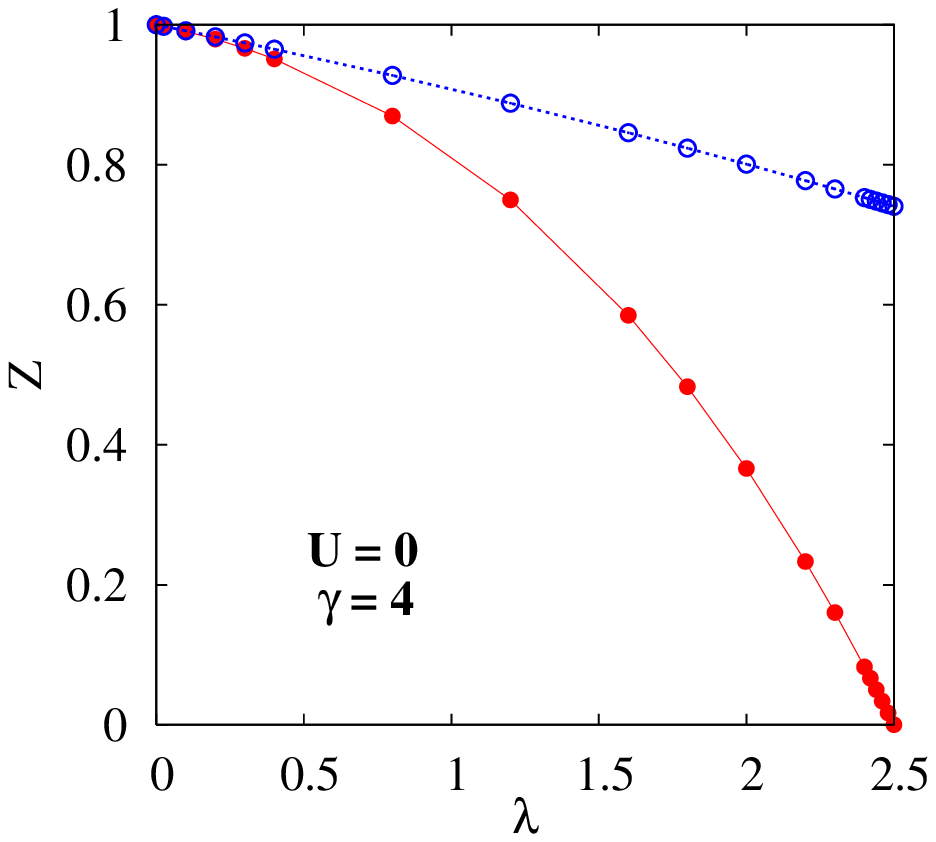}
\includegraphics[width=4.2cm]{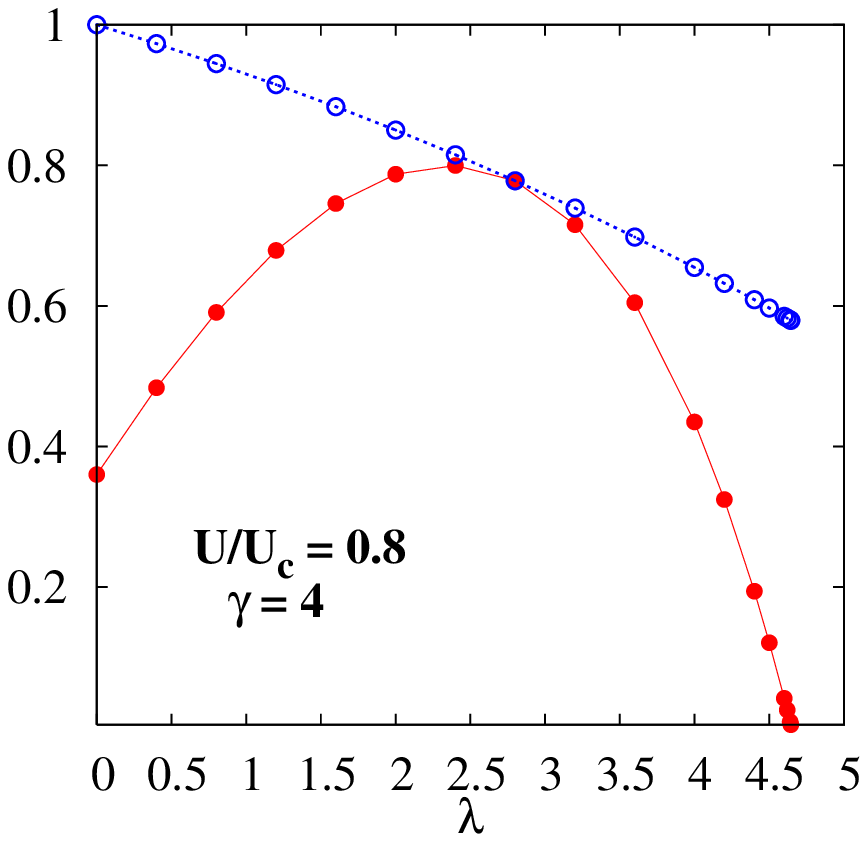}
\includegraphics[width=4.2cm]{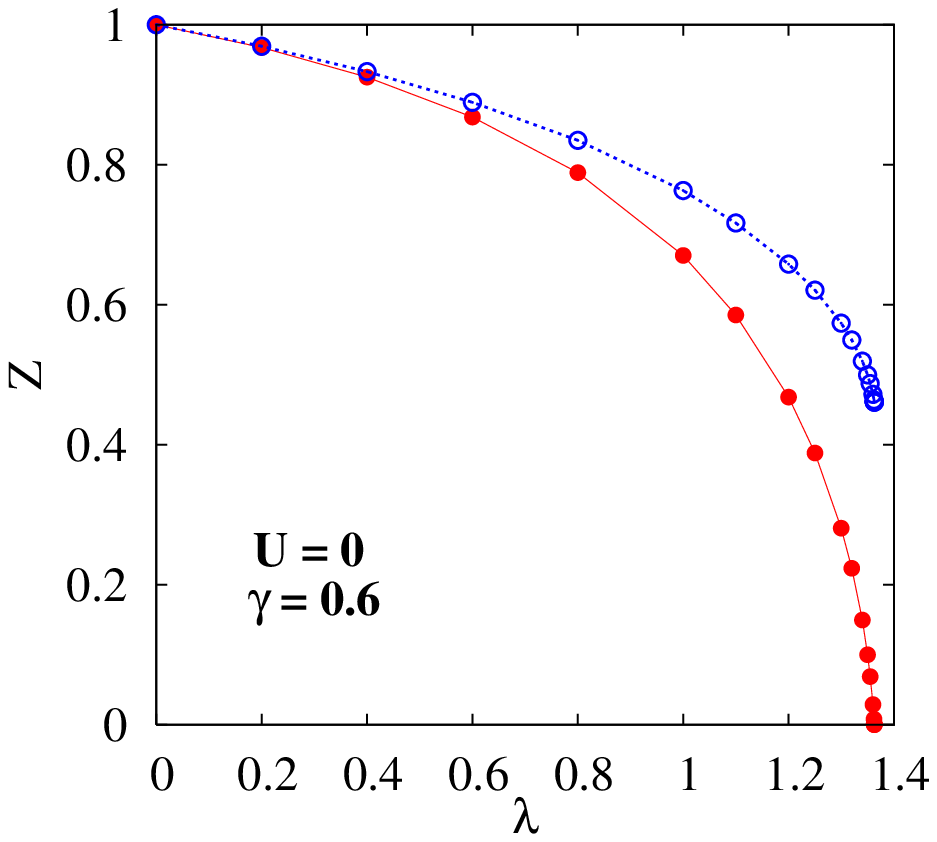}
\includegraphics[width=4.2cm]{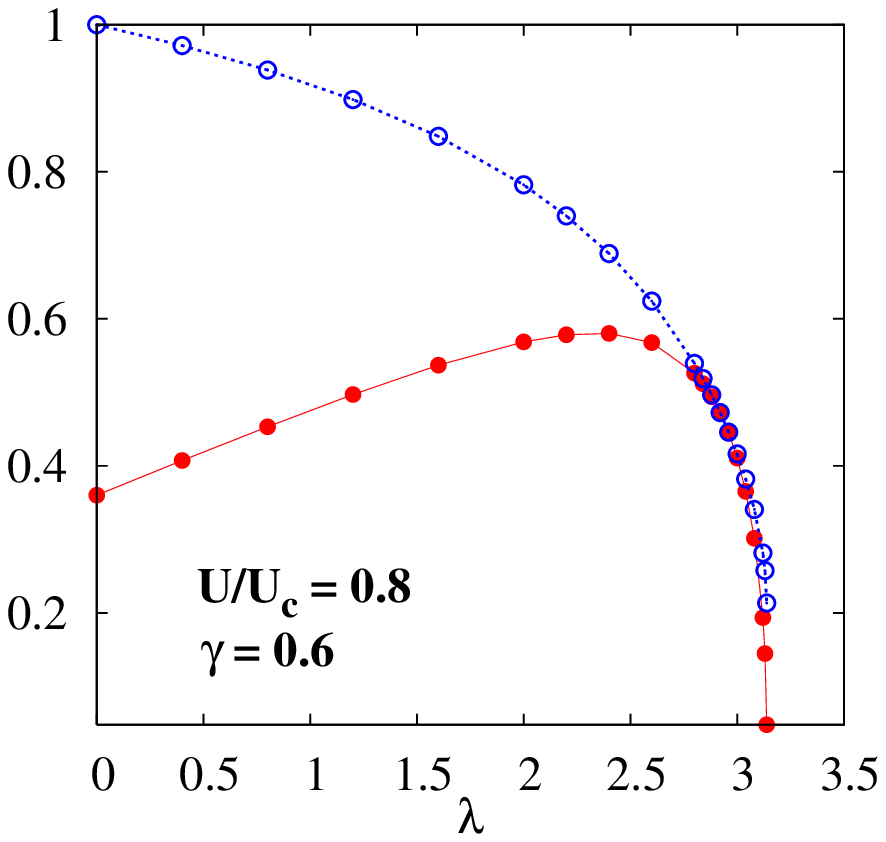}
\includegraphics[width=4.2cm]{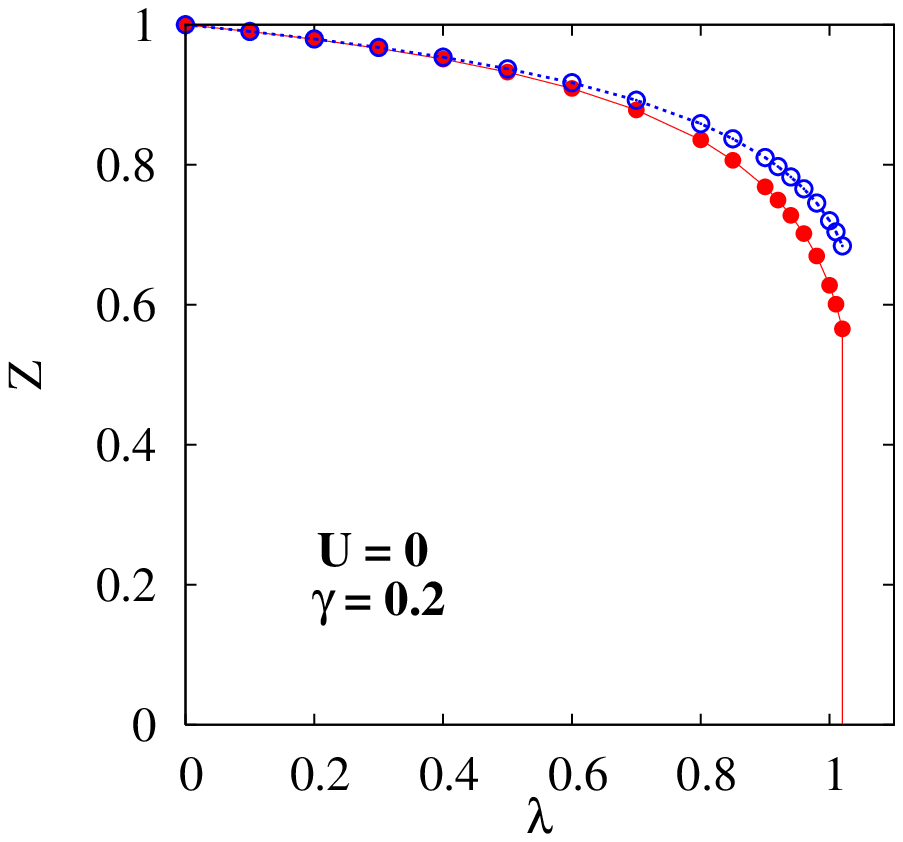}
\includegraphics[width=4.2cm]{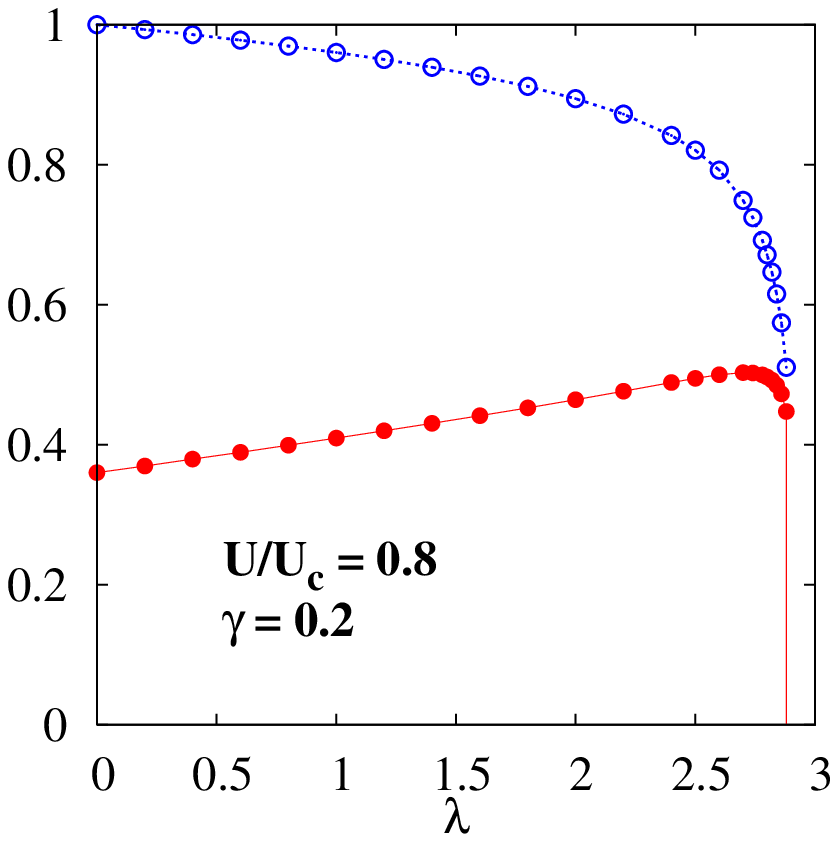}
\caption{(Color online) Typical evolution of $Z$ as a function of $\lambda$ with and without Hubbard repulsion. Parameters are $\gamma=4, 0.6,
0.2$ from top to bottom and $U=0$,
$U=0.8U_c$ in the left and right panel respectively. Open circles are the phonon contribution to the hopping renormalization
coming from $|\langle\phi_1|\phi_2\rangle|^{2}$.
}\label{fig:z}
\end{figure}
The effect of the e-ph coupling on the mass renormalization is not only comprised in the overlap
$\langle\phi_1|\phi_2\rangle$, as in the infinite-$U$ limit\cite{io_gpw}, but it is also hidden in the effective correlation
parameter $\bar{u}$. 

In the absence of $U$ it is clear from Eq. (\ref{z}) that the main correction to $Z$  for weak e-ph coupling effects comes from the overlap
$\langle\phi_1|\phi_2\rangle$, which actually results in an enhancement of the effective mass that is linear in $\lambda$, as shown in the left
panels of Fig. \ref{fig:z}; however the rapid suppression of $Z$ as the BPI is approached by increasing $\lambda$ mainly comes from the
phonon-induced attractive interaction, at least for $\gamma > 0.5$ (top and middle panels of the left column in Fig. \ref{fig:z}).
Approaching the adiabatic regime  the relevance of the effective phonon-mediated attraction is reduced, with $Z$ almost completely determined by the behavior of the overlap $\langle\phi_1|\phi_2\rangle$ as a function of $\lambda$ (bottom panel of the left column in Fig. \ref{fig:z}). Yet, the sudden decrease of the overlap drives a sharp first-order transition to the BPI for $\gamma\lesssim 0.5$.

Moving to the correlated metallic phase with $U=0.8U_c$, we find that, for all the considered values of $\gamma$, $Z$ has a non monotonic behavior. For small coupling $Z$ increases with $\lambda$, it reaches a maximum before decreasing to eventually reach the BPI. Also in this case Eq.(\ref{z}) allows to explain in a very intuitive way the observed increase of $Z$
when the e-ph coupling is turned on in a correlated regime\cite{dmft-koller,dmft-capone}.
In fact, as long as the effective interaction stays repulsive, the main
effect of $\lambda$ is to partially screen the bare Hubbard repulsion, the effectiveness of such screening being strongly
dependent on the adiabaticity regime\cite{dmft-capone} (cfr. the right panels of Fig. \ref{fig:z}); the electron motion, even if slowed down by e-ph
coupling, can take advantage of the smaller energy cost associated to double occupation, and this results in a decrease of $m^*$  with
respect to the $\lambda=0$ value. As the phonon-induced attraction overcomes the Hubbard repulsion, the two effect of phonons (localizing tendency to
self-trap and to form local pairs) are cooperative, leading to a sharp first-order transition to the BPI  for strong enough $U$ and small $\gamma$ \cite{dmft-koller}.

\begin{figure}[h]
\includegraphics[width=4.2cm]{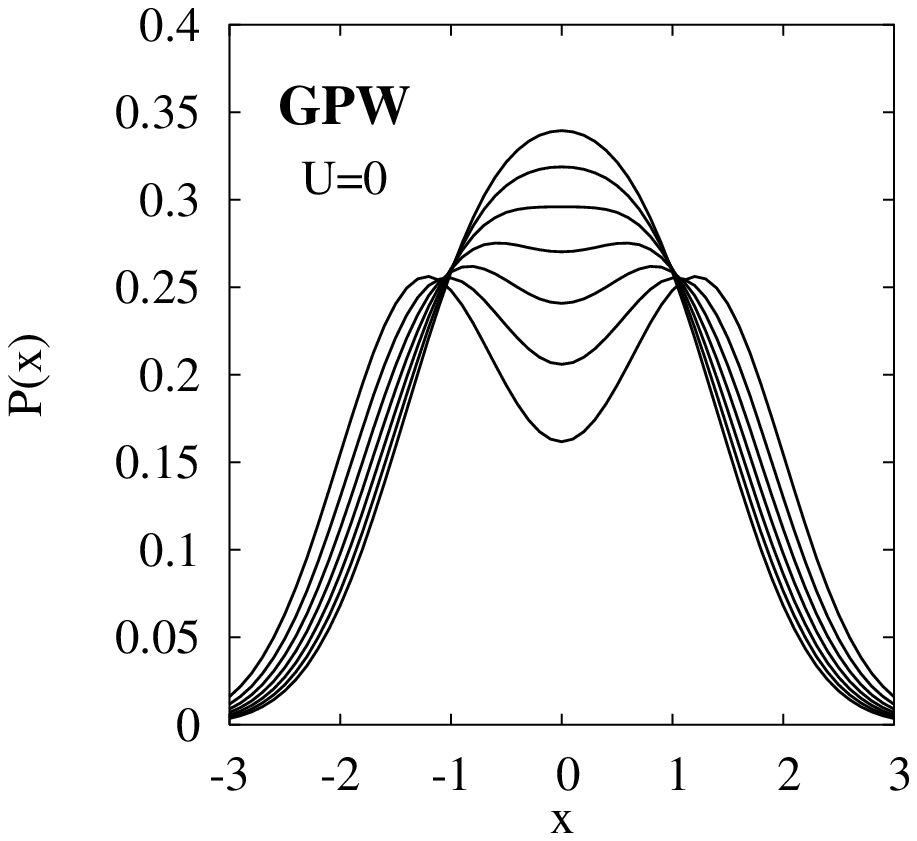}
\includegraphics[width=4.2cm]{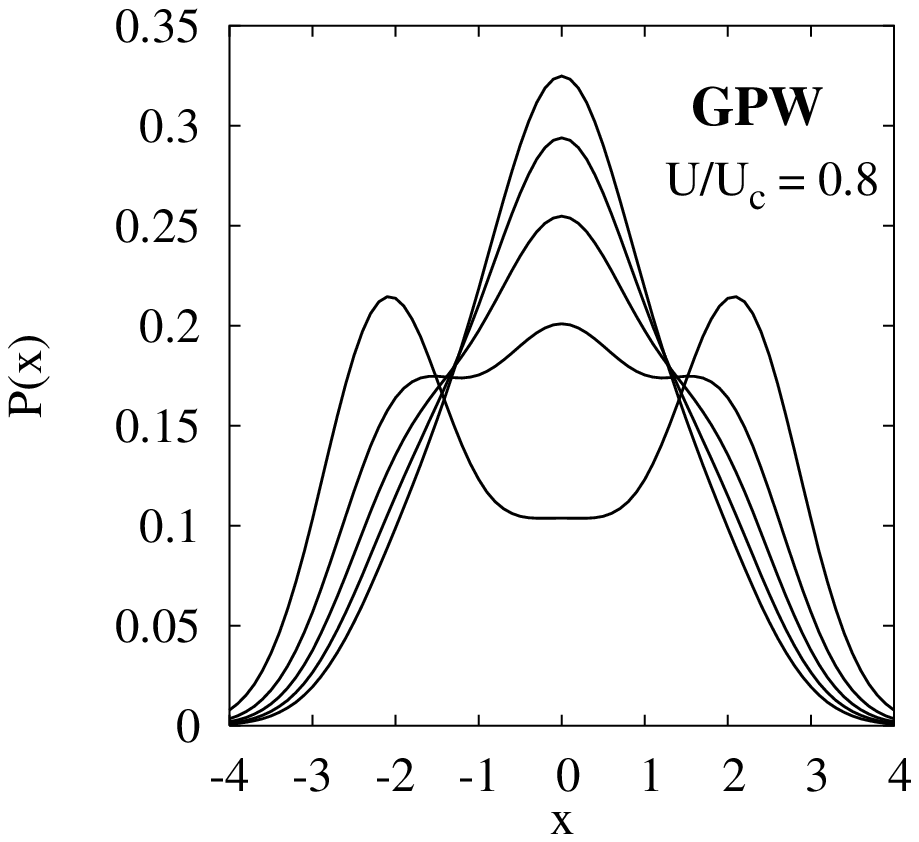}
\includegraphics[width=4.2cm]{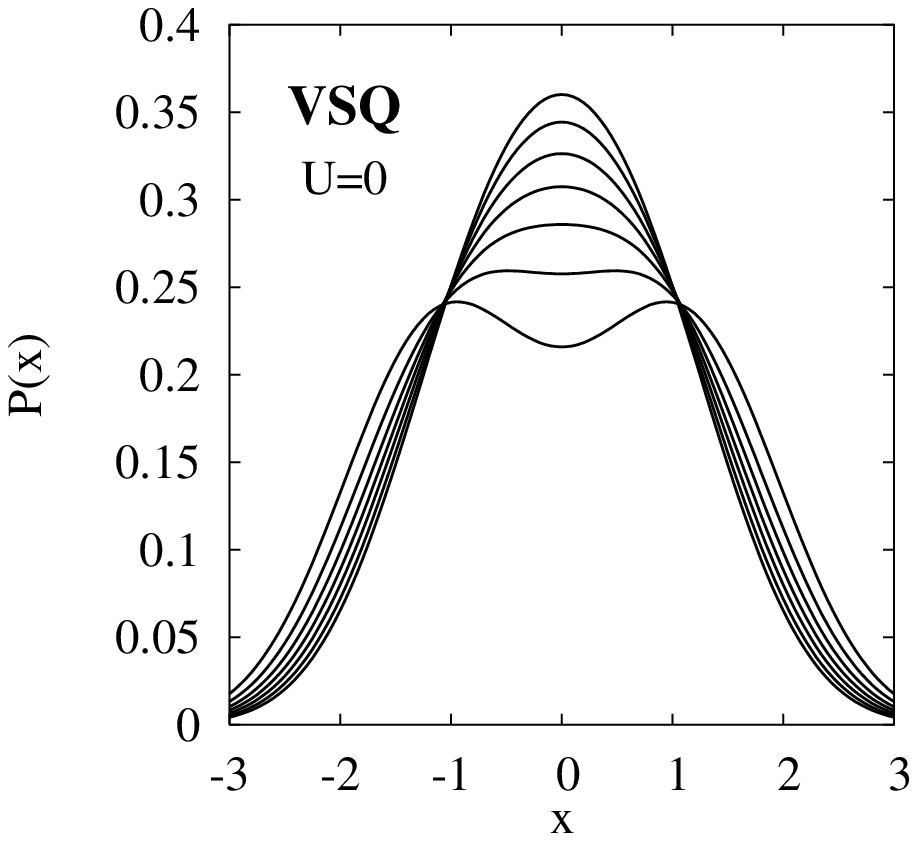}
\includegraphics[width=4.2cm]{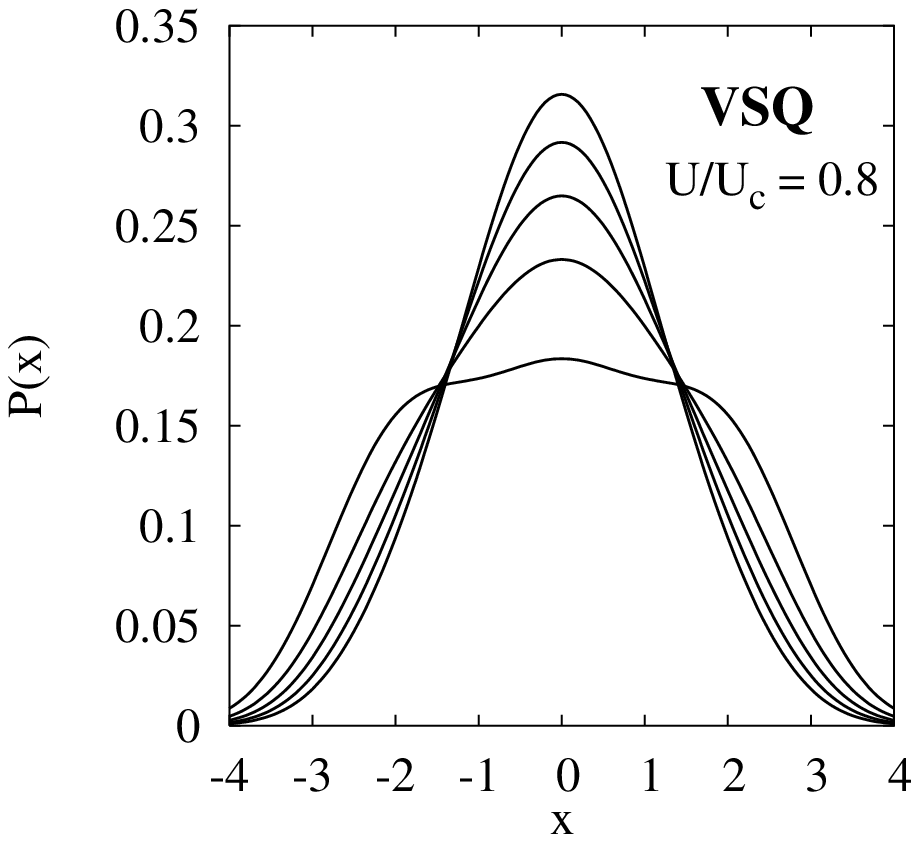}
\caption{GPW results (top panels) for $P(x)$ as a function of e-ph coupling constant at $\gamma=0.6$ in the presence of Hubbard repulsion (right panel, $U/U_c=0.8$ and $\lambda$
ranging from 2.8 to 3.12) and with $U=0$ (left panel, $\lambda$ ranging from 1 to 1.3).
The bottom panels show the lattice distribution function as obtained with VSQ for the same parameters. }\label{fig:pdf}
\end{figure}

The above discussion clearly shows that $Z$ is not the right quantity to identify the bipolaron crossover in the presence of competing interactions. While in the absence of $U$ bipolaron formation is usually associated to an abrupt enhancement of the effective mass, as shown in the left panels of Fig.~\ref{fig:z}, a more direct measure of polaronic features in the groundstate has to be used when the effective mass results from the competition between attractive and repulsive interactions. More precisely, we need a quantity that measures the entanglement between the electron motion and the lattice distortion.
A good candidate is the groundstate phonon distribution function
\begin{equation}
P(x) \equiv \langle\psi_0 \vert x\rangle\langle x\vert\psi_0\rangle = \sum_{\nu}P_{\nu}\vert\phi_{\nu}(x)\vert^2,
\end{equation} 
$\vert x\rangle$ being an eigenstate of the displacement operators, and $\vert \psi_0\rangle$ the groundstate vector.
$P(x)$ measures the effective (ground-state) displacement of the lattice.
In fact, at least in the $U=0$ limit the development of polaronic features is associated to the evolution from a monomodal shape of $P(x)$ centered around an equilibrium displacement ($x=0$ at half-filling) to a bimodal distribution with maxima displaces of $\pm x_0$, signaling the development of finite lattice distortions\cite{caponeciuchi}. Focusing on the half-filling case, $x=0$ changes from a maximum in the monomodal distribution to a minimum in the bimodal one. Thus we can associate the appearance of bipolarons in our model to a change of sign in the second derivative of $P(x)$ for $x=0$, i.e.,
\begin{eqnarray}\label{polf_cond}
4d\left[\phi_2(x=0)\frac{d^2 \phi_2}{d x^2}\bigg\vert_{x=0} +\left(\frac{d \phi_2}{d x}\right)^2_{x=0} \right] + \nonumber &&\\
2(1-2d)\,\phi_1(x=0)\frac{d^2 \phi_1}{d x^2}\bigg\vert_{x=0} \geq 0.&&
\end{eqnarray}
We notice that the evaluation of the condition (\ref{polf_cond}) is particularly sensitive to the choice of the variational projected wavefunctions.
The evolution of $P(x)$ for our Gutzwiller function is shown in Fig. \ref{fig:pdf} in comparison with the VSQ ansatz in a moderately adiabatic situation $\gamma =0.6$. 
We notice that some care has to be taken in using the condition (\ref{polf_cond}) to pinpoint the bipolaron crossover in the correlated regime (right panels), since the effect of $U$ on the $P(x)$ can result in  shoulders or even in a three-peak structure rather than the simple bimodal shape. However our criterion captures the correct order of magnitude of the critical $\lambda$ for the onset of a bipolaronic groundstate.

In the antiadiabatic limit $\gamma \to \infty$ no bipolaron crossover occurs before the system turns insulating by forming incoherent local pairs: Replacing  $\phi_\nu$ in Eq. (\ref{polf_cond}) with the proper displaced harmonic oscillators, one finds, in agreement with the DMFT of Ref. \onlinecite{caponeciuchi}, that bipolaron formation occurs at the transition line only when $\alpha^2>1/4$, which implies a very large $\lambda$ when $\gamma$ is large. However, at finite values of the adiabaticity parameter, the condition (\ref{polf_cond}) can be realized, since the lattice distortions induced by the e-ph coupling, which are roughly proportional to $1/\sqrt{\gamma}$, can push the system to polaron formation before the BPI instability takes place. Actually, for all the considered values of $\gamma$ we found bipolaron formation before the metal-BPI transition line, in a region whose size depends again on both correlation and adiabaticity parameter.
On the weakly correlated side, we notice (see Fig. \ref{fig:phasedia}) that, starting from small $\gamma$, the distance between the bipolaronic crossover and the metal-insulator transition first increases before decreasing in the antiadiabatic regime. This is in qualitative agreement with DMFT\cite{caponeciuchi}, even if for $\gamma=4$ the GPW still predicts the bipolaronic crossover before the BPI transition, in contrast with DMFT, where for the same value of $\gamma$ the opposite occurs.
 As shown in Fig. \ref{fig:phasedia}, at $\gamma=4$, the presence of $U$ slightly enlarge the region where the ground state displays polaronic features.

\section{Discussion of the results.}\label{sec:discussion}

In this section we discuss the behavior of the phonon wavefunctions, that represent the main novelty introduced by our Gutzwiller method with respect to previous schemes where the functional form was assumed {\it a priori}.

\begin{figure}[h]
\includegraphics[width=8cm]{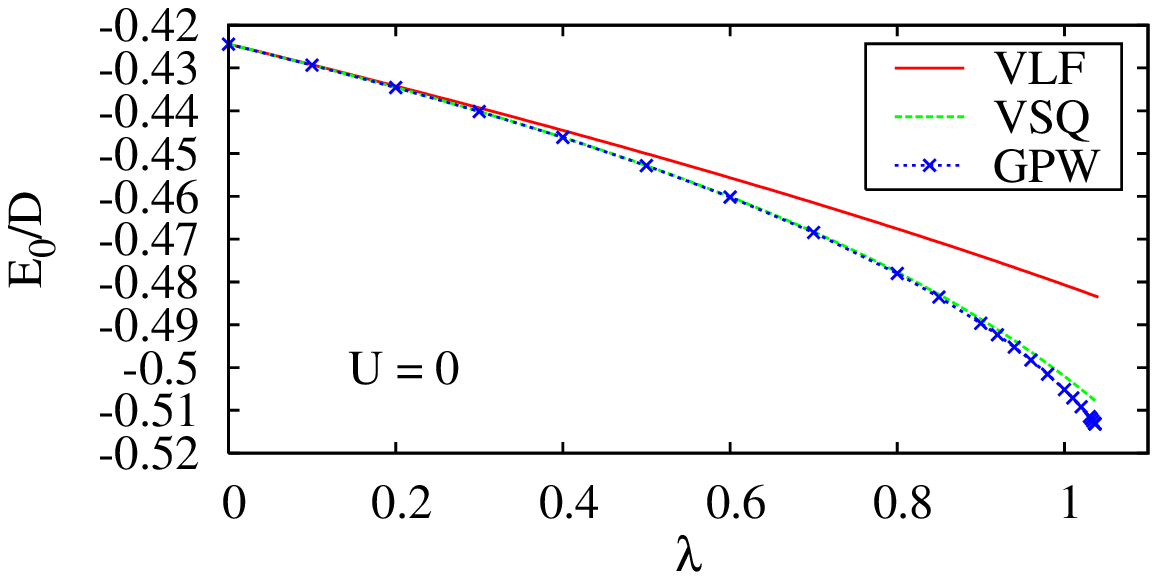}
\includegraphics[width=8cm]{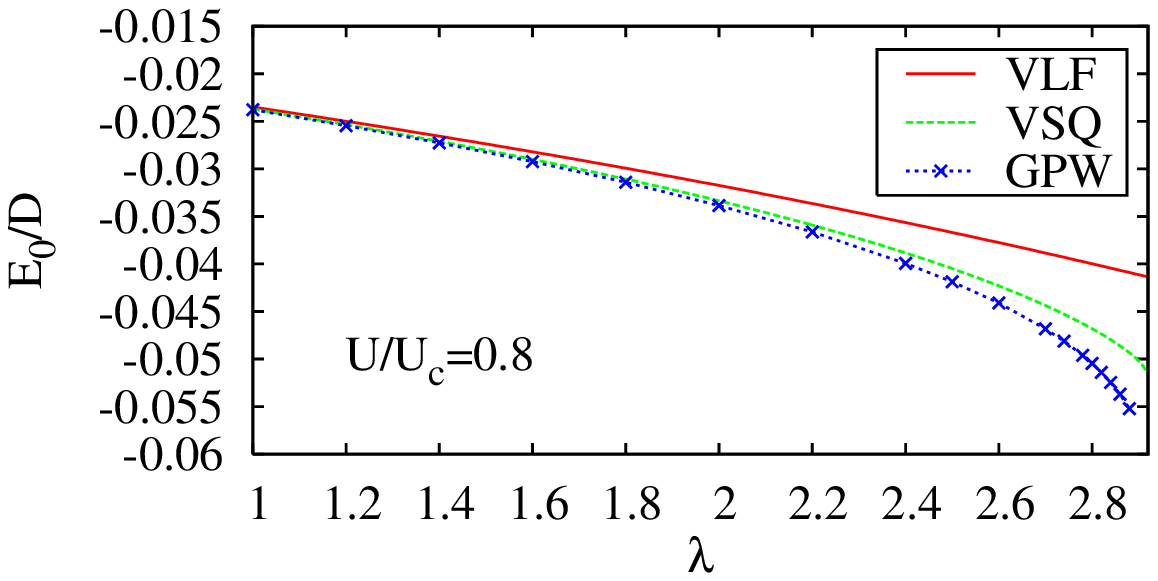}
\caption{(Color online) Comparison of the variational on-site energies obtained for the paramagnetic solution in the absence of Hubbard repulsion (upper panel) and close to the critical value $U_c=8\vert\varepsilon_0\vert$ for the purely electronic Mott transition (lower panel) for $\gamma=0.2$.}\label{fig:energies}
\end{figure}

A comparison with VLF and VSQ schemes helps us to benchmark our method. Due to the enlarged variational space, the ground-state energy of the GPW is always smaller than that given in Eqs. (\ref{vlf_energy}) and (\ref{sq_energy}). 
The gaussian ansatz for the $\phi_\nu$'s (VLF) results in a significantly higher energy, especially in the intermediate to strong coupling regime, whereas the VSQ energy stays very close to the general GPW one, even though the GPW becomes visibly more accurate than VSQ in the presence of sizeable  correlations, as shown in Fig. \ref{fig:energies}.
However, even when  GPW provides only a slight improvement on the variational energy with respect to VSQ, it gives a more accurate description of the physical properties of the lattice;  in fact, the projected phonon wavefunctions of GPW can provide a better estimates of the different energy contributions, i.e. the potential energy gain  and the kinetic energy loss due to lattice distortion, whose balance results in the overall ground-state energy.

\begin{figure}[h]
\includegraphics[width=4cm]{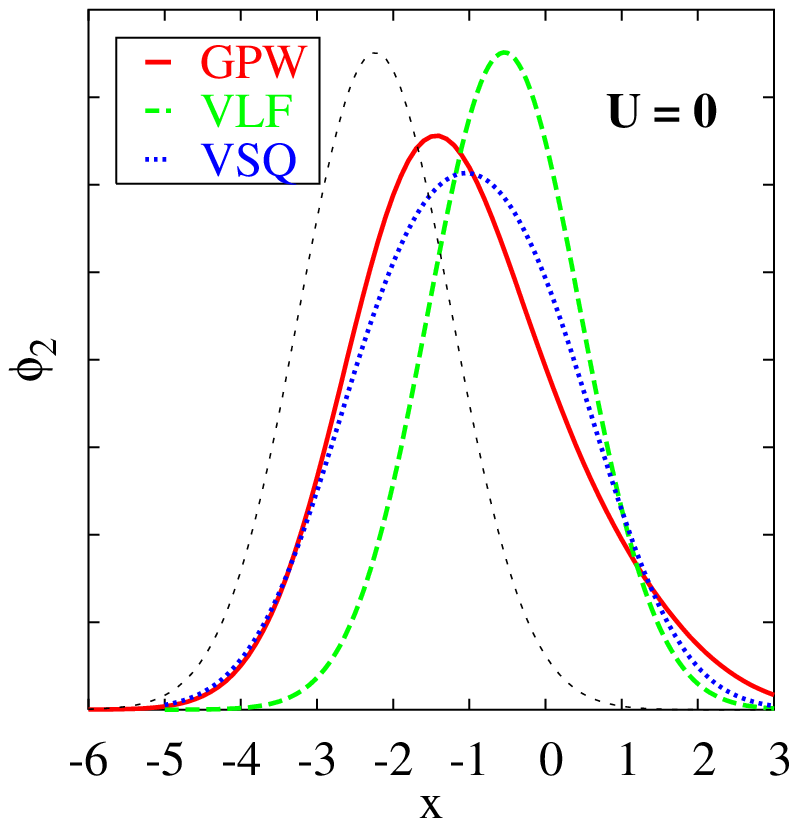}
\includegraphics[width=4cm]{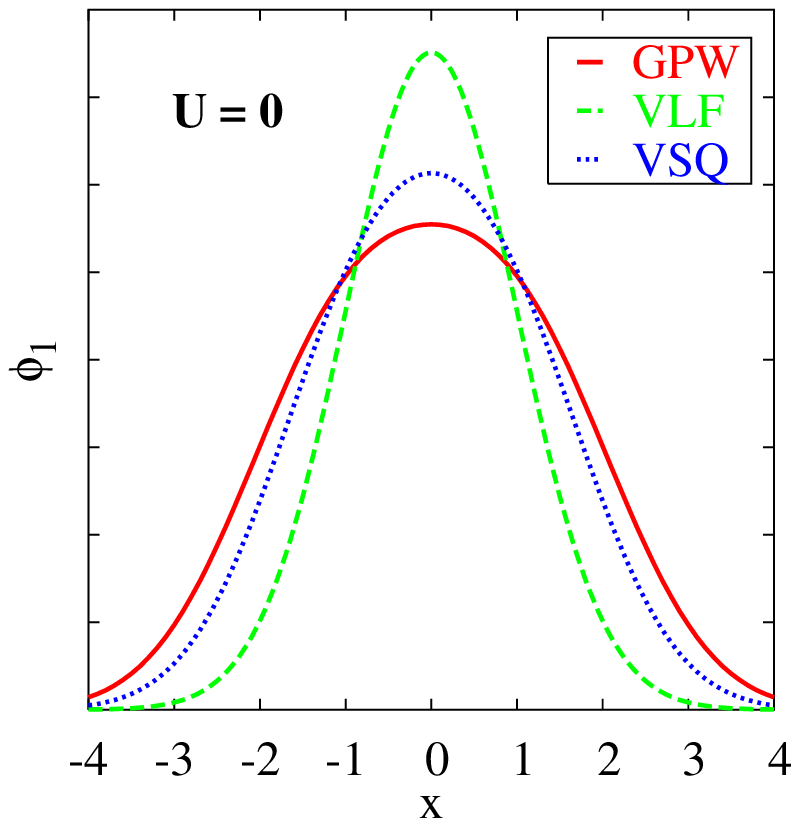}
\includegraphics[width=8.1cm]{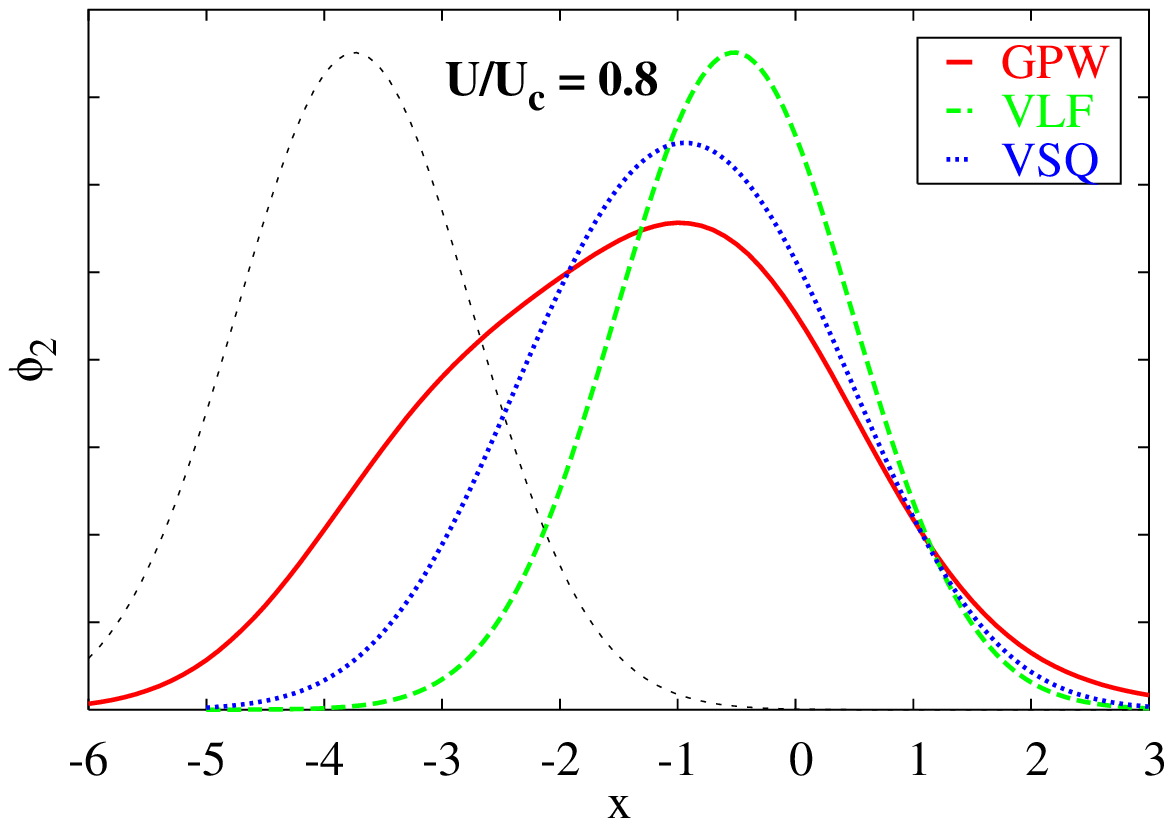}
\caption{(Color online) Comparison of the projected phonon wavefunctions for $\gamma=0.2$ as obtained with GPW and with the variational ansatzs
discussed in the text
(top panel is $U=0, \lambda=1$; bottom panel is $U= 0.8 U_c, \lambda=2.8$).
The thin dotted line is the corresponding atomic solution.}\label{fig:wavefs}
\end{figure}

To get a flavor of the improvement introduced by the wavefunction Eq. (\ref{gutzwiller}) we plot in Fig. \ref{fig:wavefs} $\phi_2$ and $\phi_1$ as obtained by means of our general equations, comparing them with the simpler variational ansatzs, VLF and VSQ. In the absence of Coulomb repulsion, the squeezed states stay relatively close to the self-consistent solution, whereas VLF underestimates the effective lattice distortion and therefore the potential energy gain. On the other hand, in the strongly-correlated regime the GPW functions are rather different from the other methods and underline the non-trivial way in which anharmonic fluctuations of the lattice states are influenced by the correlated electron motion. Even if significantly more accurate than VLF, the VSQ underestimates the anomalous lattice fluctuations that we find in GPW; since the overlap of the phonon wavefunctions related to different charge configurations is essential in capturing the way e-ph coupling renormalizes the effective hopping, this means that VSQ overestimates the kinetic energy loss.

\begin{figure}[h]
\includegraphics[width=4.27cm]{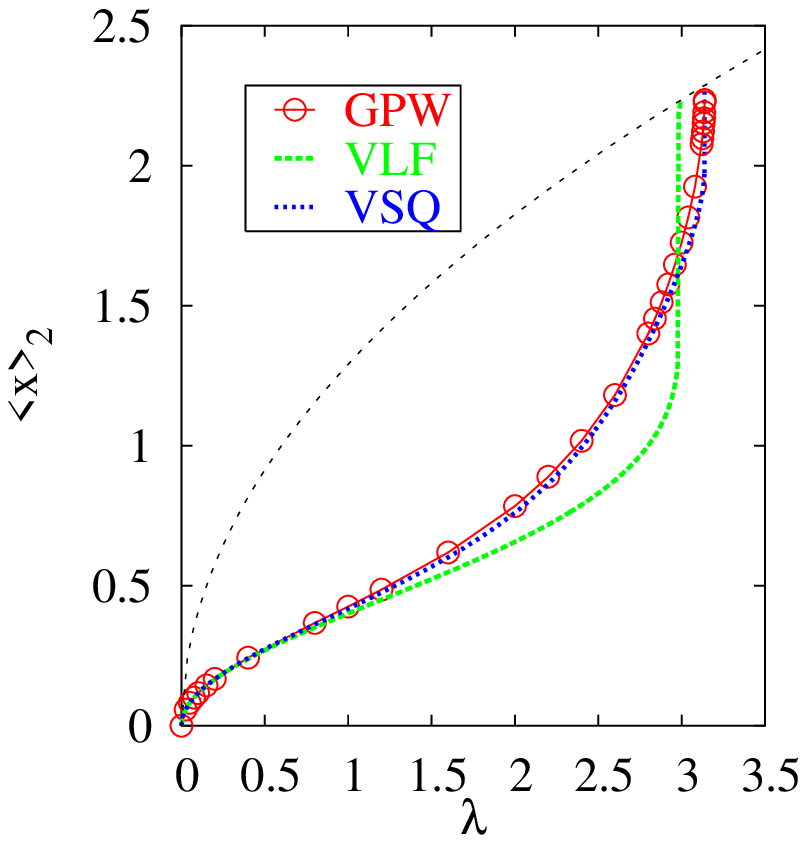}
\includegraphics[width=4.27cm]{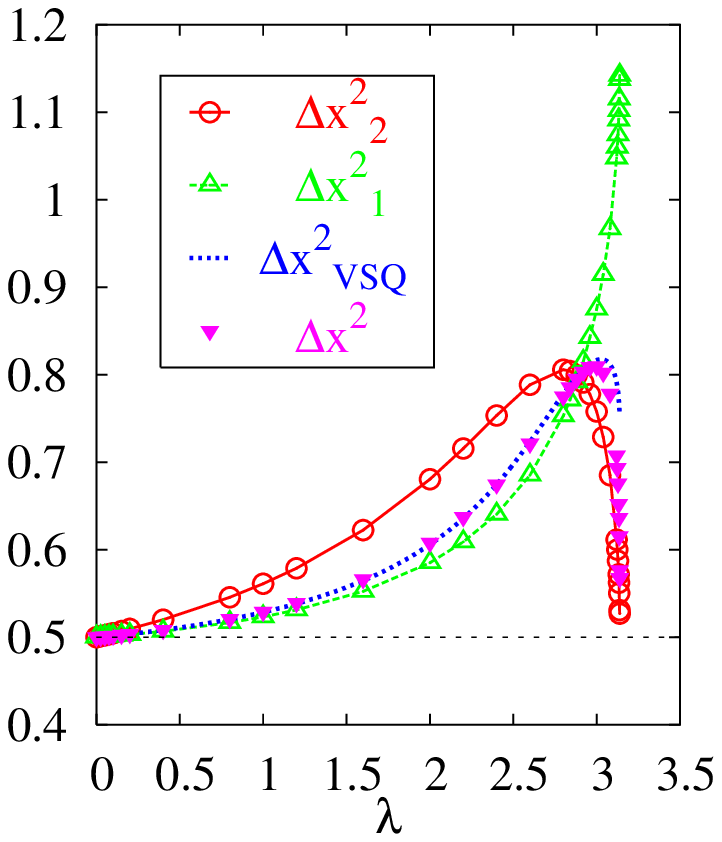}
\caption{(Color online) Comparison of the effective displacement and width of projected phonon wavefunctions in the metallic solution before the BPI is reached as obtained
with GPW and with the variational ansatzs discussed in the text (parameters are $\gamma=0.6$ and $U=0.8 U_c$). The thin dotted line is the
atomic solution. 
}\label{fig:prop-wavefs}
\end{figure}

A deeper understanding of the differences between the standard variational ansatzs and the self-consistent solution for the phonon wavefunctions can be traced out by looking at the degree of effective displacement, given by $\langle x\rangle_{\nu}$, and of the related quantum fluctuations, that can be estimated by $\Delta x^2_\nu=\langle [x_{\nu}-\langle x_{\nu}\rangle]^2\rangle$. These two quantities are, roughly speaking, related respectively to the potential energy gain and to the kinetic energy loss induced by the e-ph coupling, and their inspection will help us to identify the way in which the GPW improves with respect to the other approaches.
It is readily found that $\langle x\rangle_{\nu}=\pm\sqrt{2}\alpha f$ for
$\nu=0,2$ and  $\langle x\rangle_{1}=0$ for both VLF and VSQ ansatz, with $f$ given respectively by Eqs. (\ref{vlf_f_mf}) and
(\ref{sq_f_mf}), whereas $\Delta x^2_\nu=1/2$ in the harmonic approximation and
$\Delta x^2_\nu=1/2\tau^2$ for any $\nu$ in the squeezed phonon state approximation. In Fig. \ref{fig:prop-wavefs} we compare the
effective displacement and $\Delta x^2$ obtained within the different approaches as a function of $\lambda$ in the presence of a sizeable
$U$, namely $U=0.8 U_c$. By looking at $\langle x\rangle_{\nu}$, we find that both harmonic and squeezed ansatzs reproduce the weakly
interacting regime quite accurately, while VLF deviates in the intermediate to strongly-coupled regime, falling suddenly onto the atomic
solution (which in the present mean-field approach actually translates in the BPI). The origin of such a discrepancy is the relevance of phonon quantum fluctuations that are crucial in the crossover region and are poorly described within VLF.
However within GPW such effect is strongly dependent on the charge state associated to the different wavefunctions: the phonon wavefunctions
projected onto empty and doubly-occupied sites display a maximum in $\Delta x^2_{0(2)}$ and then tend to the atomic solutions as the BPI is
approached, whereas $\Delta x^2_1$ increases monotonically. Therefore, as anticipated, the simple two-phonon coherent state Eq. (\ref{ansatz:vsq}) reproduces only a kind of average squeezing effect; to stress this point we show an averaged $\Delta x^2$, namely $\Delta x^2=\sum_\nu P_\nu \Delta x^2_\nu$, that indeed stays very close to the VSQ solution.

In conclusion, the improved accuracy of our variational phonon wavefunctions proves to be relevant especially for intermediate-to-strong e-ph coupling and in the presence of sizeable correlation, in particular close to the BPI instability, when quantum fluctuations of the phonons play an important role. Indeed, by looking at Fig. \ref{fig:prop-wavefs},
one immediately realizes that the evaluation of the phonon distribution function strongly relies on the quality of the projected wavefunctions, and the fulfillment of  the condition (\ref{polf_cond}) can lead, as it actually does, to completely different bipolaron crossover lines for different variational ansatzs when considering adiabatic regimes. To be more explicit, we
find that the simple variational Lang-Firsov (or harmonic) ansatz never predicts the onset of a bipolaronic groundstate  for any $U$ at $\gamma<1$, being  $P(x)$ always monomodal (results not shown). On the other hand VSQ wavefunctions reproduce a bimodal $P(x)$, displayed in the bottom panels of Fig. \ref{fig:pdf}, at least at $U=0$, but at larger $\lambda$ with respect to GPW; however they completely  miss the complex structures that $P(x)$ develops in the presence of a sizeable $U$, and they do not reproduce any bipolaron crossover in the correlated regime.

\section{Conclusions}

In this work we have extended the analysis of a new type of Gutzwiller variational wavefunction introduced to
describe correlated electron systems in the presence of e-ph coupling.
In a previous paper\cite{io_gpw} we benchmarked the wavefunction  in the infinite-$U$ limit by comparing it to exact DMFT calculations, showing that polaron formation occurs smoothly for any value of the adiabaticity parameter $\gamma$ and electron density $n$, as opposed to the adiabatic first-order transition found in standard variational approaches\cite{fehske,io_vlf}.
In the infinite-$U$ limit the effect of correlation on polaronic physics can be analyzed by tuning electron filling, and we found that polaron formation is inhibited and pushed to higher, though finite, $\lambda$ when moving from low-density to half-filled system. 

Relaxing the assumption of infinite repulsion between electrons, in this paper we turned our attention to the half-filling regime and considered, together with bipolaron formation, the competition between $U$ and the phonon-mediated bipolaron instability, ruled out by an infinite $U$. We found, in excellent agreement with previous
works\cite{dmft-capone,perroni,io_vlf}, that the Mott
metal-insulator transition is always robust with respect to polaron formation and that screening of
the bare repulsion due to the coupling with the lattice is less and less effective as adiabatic
regimes are attained. On the other hand we observed that both bipolaron crossover and metal-BPI
transition depend not only on the adiabaticity parameter but also on the strength of Hubbard
repulsion. Two localizing mechanisms are involved at intermediate-to-strong e-ph coupling,
one resulting in self-trapping of electrons coupled with Holstein
phonons, and the other related to the phonon-mediated e-e attraction.
If on one hand correlation pushes to larger values of $\lambda$ both the onset of bipolaronic features and the metal-BPI transition, it influences the two processes in a different way, which depend on the
adiabaticity regime, giving rise to different scenarios. For example, when $\gamma>1$, $U$
is more effective in screening the attractive interaction than in preventing the bipolaron crossover\cite{dmft-capone}.
On the other hand at small $\gamma$ the bipolaron crossover is inhibited by $U$ faster than bipolaronic transition, and polaronic features are observed in a narrow region that rapidly shrinks with increasing $U$; at the same time a change in the order of the transition occurs depending on the strength of correlation.

A critical comparison with standard variational approaches to Holstein phonons shows that the proposed wavefunction can be viewed as a proper generalization apt to include correlation effects, underlying the importance of a reliable variational description of anharmonic fluctuations in the presence of $U$. Our wavefunction improves on previous variational approaches already in the absence of electron correlations (VLF and VSQ can indeed be obtained as restricted solutions of our method limiting the variational freedom), but it still has some limitations in the deep adiabatic limit due to the nature of our phonon wavefunctions, that are associated to the different local electronic states.
For the same reasons, our method becomes considerably more accurate in the presence of strong correlations, that make the electronic physics more local.

\begin{acknowledgements}

We are grateful with S. Ciuchi for important discussions and a careful reading of the manuscript, and with G. Sangiovanni for discussions.  We acknowledge financial support from MIUR PRIN, Prot. 2005022492.
\end{acknowledgements}


\begin{thebibliography}{99}
\bibitem{notabethe} In the infinite coordination limit $z \to \infty$ the hopping matrix element has to be scaled as $t/\sqrt{z}$ to obtain a finite average kinetic energy.
\end{thebibliography}
\end{document}